\documentclass[iop,numberedappendix,twocolappendix]{emulateapj}
\usepackage{amsmath, amsfonts, amssymb}

\shorttitle{Radio flares from GRBs}
\shortauthors{Kopa\v c et al.}

\newcommand{\nurs}{\nu_{m,\rm r}}
\newcommand{\nufs}{\nu_{m,\rm f}}
\newcommand{\fprs}{F_{\rm \nu,max,r}}
\newcommand{\fpfs}{F_{\rm \nu,max,f}}

\begin{document}

\title{Radio flares from gamma-ray bursts}

\author{D.~Kopa\v c$^1$, C.~G.~Mundell$^{1,2}$, S.~Kobayashi$^1$, F.~J.~Virgili$^1$, R.~Harrison$^3$, J.~Japelj$^4$, C.~Guidorzi$^5$, A.~Melandri$^6$, A.~Gomboc$^4$}
\affil{$^1$Astrophysics Research Institute, Liverpool John Moores University, Liverpool, L3 5RF, UK; D.Kopac@ljmu.ac.uk \\ 
$^2$Department of Physics, University of Bath, Claverton Down, Bath, BA2 7AY, UK\\
$^3$ Department of Astrophysics, School of Physics and Astronomy, Tel Aviv University, 69978 Tel Aviv, Israel \\
$^4$Faculty of Mathematics and Physics, University of Ljubljana, Jadranska 19, 1000 Ljubljana, Slovenia \\
$^5$Department of Physics and Earth Sciences, University of Ferrara, Via Saragat, 1, 44122 Ferrara, Italy \\
$^6$INAF/Brera Astronomical Observatory, via Bianchi 46, 23807, Merate (LC), Italy \\
}

\begin{abstract}
We present predictions of centimeter and millimeter radio emission from reverse shocks in the early afterglows of gamma-ray bursts with the goal of determining their detectability with current and future radio facilities. Using a range of GRB properties, such as peak optical brightness and time, isotropic equivalent gamma-ray energy and redshift, we simulate radio light curves in a framework generalized for any circumburst medium structure and including a parametrization of the shell thickness regime that is more realistic than the simple assumption of thick- or thin-shell approximations. Building on earlier work by \citet{mundell07} and \citet{melandri10} in which the typical frequency of the reverse shock was suggested to lie at radio, rather than optical wavelengths at early times, we show that the brightest and most distinct reverse-shock radio signatures are detectable up to 0.1$-$1 day after the burst, emphasizing the need for rapid radio follow-up. Detection is easier for bursts with later optical peaks, high isotropic energies, lower circumburst medium densities, and at observing frequencies that are less prone to synchrotron self-absorption effects - typically above a few GHz. Given recent detections of polarized prompt gamma-ray and optical reverse-shock emission, we suggest that detection of polarized radio/mm emission will unambiguously confirm the presence of low-frequency reverse shocks at early time. 
\end{abstract}


\section{Introduction}

In the standard fireball model of gamma-ray bursts (GRBs), internal shocks in the expanding flow produce the prompt $\gamma$-ray emission that characterizes a GRB. As the expanding ejecta collide with the surrounding circumburst medium, a fading afterglow is produced, which is comprised of two components: a forward shock (FS) that propagates outwards into the ambient medium and a reverse shock (RS) that travels backwards into the on-coming flow \citep{meszaros94,meszaros94b}. It was expected that  bright optical flashes, produced by the RS similar to that found in GRB~990123 \citep{akerlof99,sari99b,kobayashisari00,kobayashi00}, would be common in the early afterglows of GRBs.

Despite ten years of accurate GRB localizations disseminated automatically in real-time from \textit{Swift} \citep{gehrels04} and rapid, ground-based follow-up by autonomous robotic optical telescopes (e.g., \citealt{monfardini06, guidorzi11, virgili13}), only a small fraction ($\sim$5\%) of early-time optical light curves show clear evidence of optical RS emission \citep{meszaros99,sari99,kobayashi00,roming06,gomboc08,kopac13,japelj14}.  

\citet{mundell07} suggested that this lack of bright optical RS emission may be explained if the typical synchrotron frequency of the RS already lies at radio frequencies at early time. \citet{melandri10} extended this low frequency scenario to a sample of 19 GRBs with well-sampled optical light curves, each with a single optical peak that was consistent with the typical frequency of the FS lying close to the optical band. They produced model radio light curves for FS and RS emission, including a simple parametrization of synchrotron self-absorption at early time. 

Radio interferometers, such as the Jansky Very Large Array (JVLA), the Atacama Large Millimeter/submillimeter Array (ALMA) and pathfinders for the Square Kilometer Array, will come on line or have been significantly upgraded to provide unprecedented sensitivity, providing mJy and $\mu$Jy level observations in relatively short integration times and at ever-shorter response times (e.g., \citealt{chandra12, laskar13, laskar14}). Ultimately, sensitive radio surveys of the transient sky will become routine with future facilities such as the Square Kilometer Array (SKA). Estimates for radio GRB event rates are beginning to emerge but these currently neglect RS emission, focusing instead on FS emission, for which radio light curves have traditionally been better characterized \citep{ghirlanda13,burlon15,metzger15}. Motivated by this, and following \citet{mundell07} and \citet{melandri10}, we provide new theoretical predictions of GRB radio light curves including {\em RS emission} within an updated low-frequency model framework, incorporating corrections from hydrodynamical simulations and a generalized circumburst medium. The latter addition stems from the practice of analyzing GRBs in a constant ISM ($k=0$) and wind-type medium ($k=2$), but also from modeling of GRBs with structured optical emission that imply an intermediate stratification ($k\sim1$, \citealt{liang13, yi13}) and detailed hydrodynamical light curve modeling (e.g. \citealt{decolle12}).

In Section 2 we present the updated low-frequency model formalism, in Section 3 we detail our Monte Carlo simulations, in Section 4 we present our results, in Section 5 we discuss the implications of the simulations, and in Section 6 we highlight the conclusions.

\section{Low-frequency model framework}
\label{sect:model}

For GRBs with well-sampled early optical afterglows we use a set of simple assumptions about the relationship between the RS and FS emission at the deceleration time to construct predicted low-frequency light curves similar to \citet{melandri10} but generalized for any circumburst medium structure. The subset of bursts with a single peak in their optical light curves are of particular interest because they provide information about the time of deceleration of the fireball, if we assume $t_{\rm p,opt} \sim t_{\rm dec}$\footnote{Although this assumption may not hold for bursts with late peaks (e.g. \citealt{guidorzi14}), it is still reasonable for many observed bursts \citep{hascoet14}.}. 

Using the deceleration time estimated from the optical light curve and the high-energy properties of the burst we estimate the Lorentz factor $\Gamma_0$ for a general stratified medium, where the circumburst medium density is given by $n = AR^{-k}$ \citep{chevalier99,chevalier00} and where $R$ is a distance from the progenitor star. We adopt the formalism of \citet{yi13} where $n = n_0 (R/R_0)^{-k}$, with typical values of $n_0 \sim 1 \rm ~cm^{-3}$ and $R_0 \sim 10^{17}$ cm. The Lorentz factor is approximated by:
\begin{align}
\label{eq:gamma}
\Gamma_0 \sim \, & C_t^{(3-k)/(8-2k)} \times \notag \\
& \left[ \frac{(3-k)E_{\gamma,\rm iso}}{4\pi \eta n_0 R_0^k m_p c^2} \left( \frac{1+z}{c\,t_{\rm p,opt}}\right)^{(3-k)} \right]^{1/(8-2k)} \,,
\end{align}
where $c$ is the speed of light, $m_p$ is the proton mass, $\eta$ is the radiative efficiency, $E_{\gamma,\rm iso}$ is the isotropic equivalent energy of prompt gamma-ray emission, and $C_t \propto \xi _0 ^{-2}$ is a numerical factor that corrects for the shell thickness regime, as determined from numerical hydrodynamic simulations (\citealt{harrison13}; 2015 in prep). 

At the deceleration time, the RS peak frequency and peak spectral flux density are estimated by:
\begin{equation}
\label{eq:nu_rs}
\nurs \sim C_m \Gamma_0^{-2} \nufs  
\end{equation}

\begin{equation}
\label{eq:fp_rs} 
\fprs \sim C_F \Gamma_0 \fpfs,
\end{equation}
where $C_m(\xi _0, k)$ and $C_F (\xi _0, k)$ are numerical factors that correct for the shell thickness regime for $k=[0, 1, 2]$ (\citealt{harrison13}; 2015 in prep), and `$\rm r$' and `$\rm f$' designate reverse and forward shock, respectively \citep{sari99,kobayashi03,zhang03}. Here we assume that $\epsilon_{B,\rm r} = \epsilon_{B,\rm f}$, where $\epsilon _B$ is the ratio between the magnetic energy density and the internal energy density. 

The parameter $\xi_0$ describes the shell thickness regime, which strongly affects the properties of the RS emission \citep{nakar04, harrison13}. Determined as $\xi _0 = (l/\Delta_0)^{1/2}\Gamma_0^{-(4-k)/(3-k)}$, where $l = [(3-k)E/(4 \pi n_0 R_0^k m_p c^2)]^{1/(3-k)}$ is the Sedov length and $\Delta_0 \sim c \, T_{90}/(1+z)$ is the shell width estimate \citep{sari95,kobayashi97}, $\xi _0 \ll 1$ and $\xi _0 \gg 1$ correspond to the relativistic (thick-shell) and Newtonian (thin-shell) regime, respectively. In order to allow for a variety of shell thicknesses, we approximate the observed $\xi _0$ distribution using:
\begin{equation}
\label{eq:xi0}
\xi _0 ^2 \sim C \times (t_{\rm p,opt}/T_{90}-1)\,,
\end{equation}
with the value of constant $C$ estimated from hydrodynamical simulations such that $C \sim [5, 20, 10]$ for $k=[0, 1, 2]$, respectively (\citealt{harrison13}; 2015 in prep). The estimate of $\xi _0$ depends on the assumption that shell width can be approximated by $T_{90}$; this may not be accurate in some cases (for more discussion on this, see \citealt{kobayashi00} and \citealt{nakar04}).

\begin{figure}[!h]
\begin{center}
\includegraphics[width=1\linewidth]{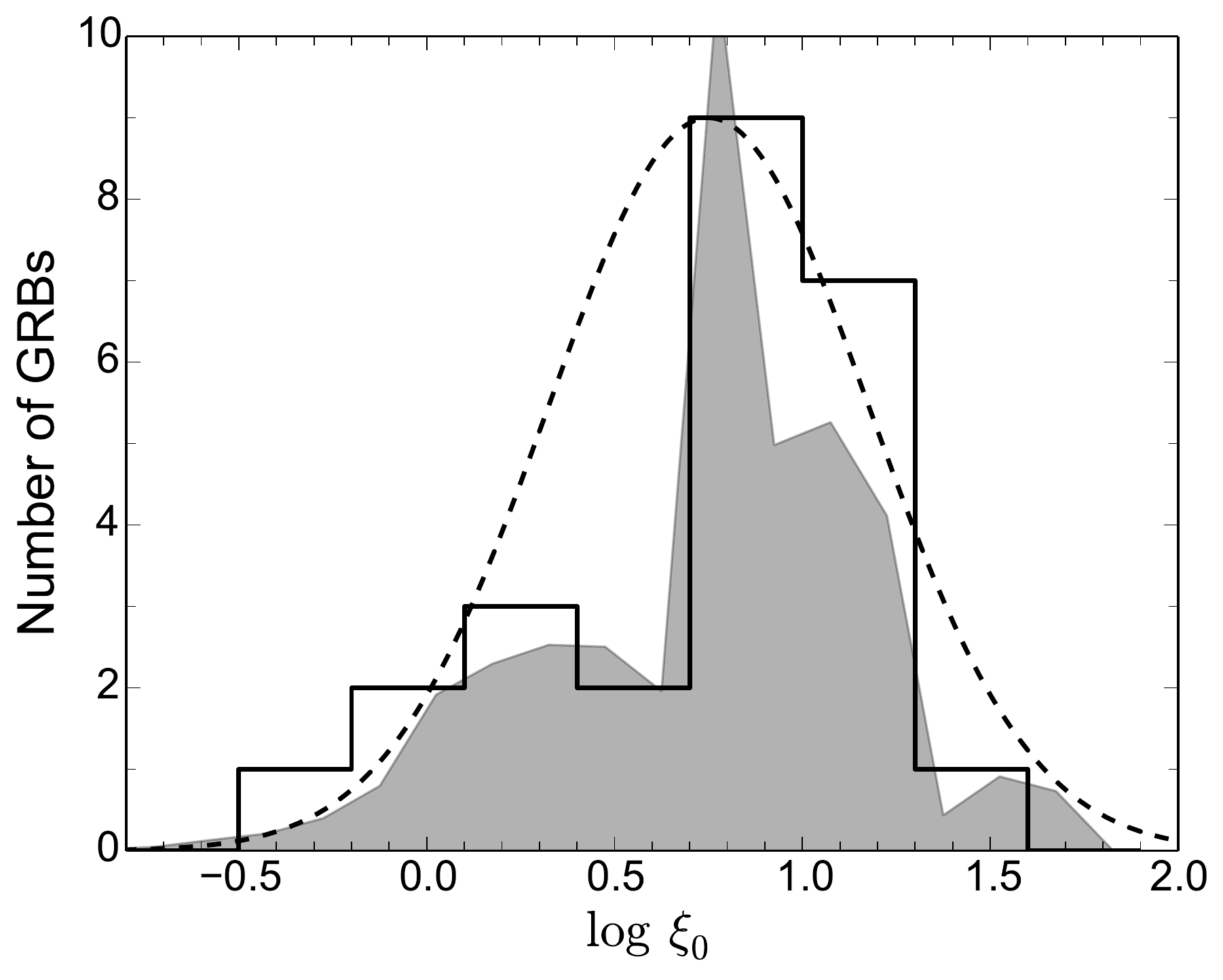} 
\caption{\label{fig:xi} Distribution of $\xi _0$ for $k=1$ case, as estimated from $25$ GRBs with clearly detected optical peaks (solid black line). Gray histogram represents the same distribution, but obtained by taking into account also the uncertainty on $t_{\rm p,opt}$ and $T_{90}$. Dashed black line is the normal distribution of $\xi _0$ values, with mean $\log \xi _0 \sim 0.75$ and $\sigma \sim 0.42$.}
\end{center}
\end{figure}

Using Equation (\ref{eq:xi0}) we estimate $\xi _0$ for $25$ GRBs with optical peaks detected from the literature \citep{rykoff09, melandri10, liang13, yi13, panaitescu13, hascoet14}. The resultant distribution for $k=1$ case is plotted in Figure \ref{fig:xi}. To include $\xi _0$ in our simulations, we approximated the distributions for each $k$ with associated log-normal distributions. Although these should not be considered as a robust fit due to the sparsity of $\xi _0$ values, they resemble well the parameter space. Using log-normal distributions, we generate random values of $\xi _0$ to calculate the numerical constants $C_m$, $C_F$, and $C_t$ for $k=[0,1,2]$ (as determined by \citealt{harrison13}; 2015 in prep), which are used in generating the low-frequency light curves.

Using Equations (\ref{eq:gamma})-(\ref{eq:xi0}), we construct our low-frequency model light curves. Light curve scalings in a generalized stratified medium used below are based on the derivations of \citet{yi13} and \citet{vanderhorst14}. Both RS and FS time dependencies change according to spectral regime and RS shell type, but we initially consider the case where the RS shell crossing has already occurred and the scalings are insensitive to RS shell thickness.

\subsection{Forward-shock emission}

For a slow-cooling synchrotron spectrum \citep{sari98}, which is expected at long wavelengths under the assumption that $\nu_{\rm radio,obs} < \nufs \sim \nu_{\rm opt}$ at $t_{\rm p,opt}$, the FS peaks at radio frequencies at $t_{\rm p,radio,f} \sim (\nu_{\rm opt}/\nu_{\rm radio,obs})^{2/3}t_{\rm p,opt}$ with a peak flux of $\fpfs \sim (t_{\rm p,radio,f}/t_{\rm p,opt})^{-k/(2(4-k))} F_{\rm p,opt}$, reflecting the evolution of $\fpfs$ with time for cases where $k \neq 0$ (e.g. \citealt{granotsari02}). Here, $\nu_{\rm radio,obs}$ and $\nu_{\rm opt}$ are the observed radio and optical frequencies, respectively. In this spectral regime, the light curve scales as $F \propto t^{(2-k)/(4-k)}$ and $F \propto t^{(-12p+3kp-5k+12)/4(4-k)}$ \citep{vanderhorst14}, before and after the peak respectively. 

Here we note that a correction factor $\chi = \nufs/\nu_{\rm opt}$ is introduced in cases where $\nufs$ is significantly below the optical band, modifying the FS peak time by a factor of $\chi^{2/3}$ and the FS peak flux by $\chi^{-(p-1)/2}$ \citep{melandri10}. Changes to RS light curves due to factor $\chi$ are propagated through Equations (\ref{eq:nu_rs}) and (\ref{eq:fp_rs}). For simplicity, the common assumption to detect bright afterglow emission is $\chi = 1$ (i.e. $\nufs = \nu_{\rm opt}$), but we discuss changes to light curves if $\chi = 0.1$ or $\chi = 0.01$ in Section \ref{sect:optprop}. 

\subsection{Reverse-shock emission}

Using Equations (\ref{eq:nu_rs}) and (\ref{eq:fp_rs}), we can estimate the peak frequency $\nurs$ and peak flux density $\fprs$ of the RS. After the RS shell crossing, $\nurs$ scales as $t^{(14k-73)/12(4-k)}$, causing the peak of the RS light curve to occur at radio frequencies at $t _{\rm p,radio,r} \sim (\nurs/\nu_{\rm radio,obs})^{-12(4-k)/(14k-73)}t_{\rm p,opt}$ \citep{sari99b, kobayashi00, vanderhorst14}. The peak flux of the RS decays after the RS shell crossing, and, taking into account Equation (\ref{eq:fp_rs}), we can approximate the RS flux at the observed frequency as:
\begin{align}
\label{eq:fp_rs_scaling}
F_{\rm \nu,r} & \sim C_F \Gamma_0 \fpfs \left( \frac{\rm t_{p,radio,r}}{t_{\rm p,opt}} \right) ^{(10k-47)/(12(4-k))} \notag \\
& \times \left\{
\begin{array}{l@{\quad}l}
(\nu_{\rm radio,obs}/\nurs)^{1/3} & \nu_{\rm radio,obs} < \nurs \\
(\nu_{\rm radio,obs}/\nurs)^{-(p-1)/2} & \nu_{\rm radio,obs} > \nurs \,, \\
\end{array}
\right. 
\end{align}
with the latter half compensating for the possible difference between the observed radio frequency and the peak frequency given by Equation (\ref{eq:nu_rs}) (e.g. \citealt{harrison13}). Finally, the light curve in the slow-cooling regime scales as $F \propto t^{-(17 - 4k)/(9(4 - k))}$ and $F \propto t ^{(-73p + 14kp + 6k - 21)/24(4-k)}$ \citep{vanderhorst14}, before and after the peak, respectively.

\subsection{Self-absorption approximation}

Synchrotron self-absorption (SSA) becomes important at low frequencies. Below the characteristic self-absorption frequency, $\nu _a$, which is determined by the condition that the optical depth is unity, the system becomes optically thick and the afterglow flux can be reduced significantly \citep{granot99}. The exact value of $\nu _a$ depends on various microphysical parameters, which are often vaguely determined. The common alternative method used to account for SSA is to approximate the SSA flux limit with a simple black body limit with the FS or RS effective temperature \citep{sari99, kobayashisari00}. It has been shown that these two methods do not differ significantly \citep{shen09}, and the latter is also more convenient within the scope of our paper, as our low-frequency model is based on scaling of observed properties rather than using a particular model-dependent formulation. At the optical peak time, the SSA flux limit in the RS region is approximated by:
\begin{equation}
\label{eq:ssa}
F_{\rm BB} \sim \pi \nu_{\rm radio,obs}^2 \epsilon_e m_p \Gamma_0 (1+z) \left(\frac{R_\perp}{D_L}\right)^2
\end{equation}
where $R_\perp \sim 2 \Gamma_0 c t_{\rm p,opt}$ is the observed size of the fireball and $D_L$ is the luminosity distance. The SSA flux limit in the FS region is larger by a factor of $\Gamma_0$ due to the higher FS temperature. Initially, the limit in the RS (FS) region scales as $F \propto t^{(5-k)/(3(4-k))}$ ($F \propto t^{2/(4-k)}$) and then as $F \propto t^{(113-22k)/(24(4-k))}$ ($F \propto t^{(20-3k)/(4(4-k))}$) after the $\nurs$ ($\nufs$) crossing \citep{sari99b, vanderhorst14}. We note that the SSA limit will mostly affect the RS emission at early times, but could also affect the FS emission at very low frequencies, however not as significantly due to a factor of $\Gamma _0$ larger limit. 

\section{Simulations}
\label{sect:simulation}

In order to examine the parameter space of possible low-frequency light curves we perform a series of Monte Carlo simulations using the model framework of Section \ref{sect:model}. Each simulation creates $1000$ light curves with the following input:
\begin{itemize}
\item{Optical peak time $t_{\rm p,opt}$: either $200\,\mathrm{s}$ (`early') or $1000\,\mathrm{s}$ (`late').}
\item{Optical peak magnitude ($R$ band): either $15\,\mathrm{mag}$ (`bright') or $18\,\mathrm{mag}$ (`dim').}
\item{Observed radio/mm frequency $\nu _{\rm radio,obs}$: $1.4$ GHz, $10$ GHz or $100$ GHz.}
\item{Circumburst medium structure $k$: 0, 1, or 2.}
\item{Circumburst medium density $n_0$: $1\,\mathrm{cm^{-3}}$. We qualitatively discuss changes to this parameter in Section \ref{sect:ism}.}
\item{Redshift: drawn from observed normal distribution $z = 1.84 \pm 0.16$ \citep{salvaterra12}.}
\item{$E_{\rm \gamma,iso}$: drawn from observed log-normal distribution $\log E_{\rm \gamma,iso} = 52.96 \pm 0.79$ \citep{melandri14}.}
\item{Shell thickness: Drawn from observed log-normal distribution (see Figure \ref{fig:xi}) $\log \xi_0 \sim [0.45, 0.75, 0.60] \pm 0.42$ (for $k=[0,1,2]$). Due to sparse $\xi _0$ distribution, we also tested our simulation using a uniform distribution and the results did not change}.
\item{Microphysical parameters: $p \sim 2.36 \pm 0.59$ (normal dist., \citealt{curran10}), $ \epsilon _e \in 0.01-0.5$ (uniform dist., \citealt{santana14}), $\eta = 0.2$ (fixed, \citealt{santana14}). Parameter $\epsilon _\mathrm{B}$ is not explicitly used in our formulation because we assume $\epsilon_{B,\rm r} = \epsilon_{B,\rm f}$ and use FS properties obtained from optical observations to normalize FS emission, while the RS properties are determined through the shock jump conditions \citep{kobayashi03}.}
\end{itemize}

\section{Results}

\begin{figure*}[!ht]
\begin{center}
\includegraphics[width=0.82\linewidth]{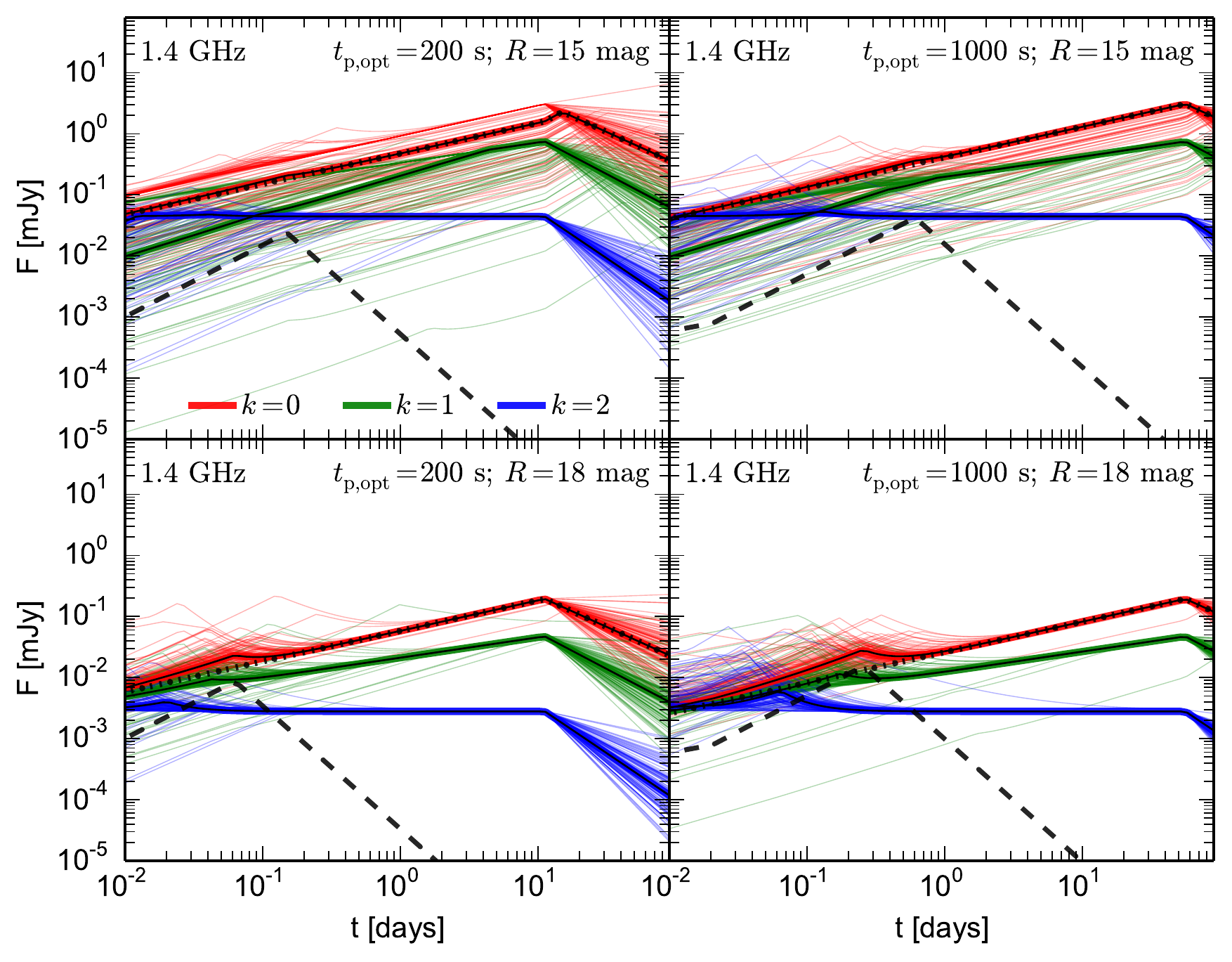}  
\caption{Predicted light curves for 1.4 GHz. Thick lines represent the median light curves, while thin lines represent simulated light curves (Section \ref{sect:simulation}; we randomly chose $100$ out of $1000$ simulated light curves for clarity). Colors indicate different circumburst medium stratification $k$. Four panels represent various peak times $t_{\rm p,opt}$ and peak magnitudes $R$. Also shown for $k=0$ cases are the relative contribution of the RS corrected for corresponding SSA limit (dashed black line), and of the FS corrected for corresponding SSA limit (dotted black line).}
\label{k0lc}
\end{center}
\end{figure*}

\begin{figure*}[!ht]
\begin{center}
\includegraphics[width=0.82\linewidth]{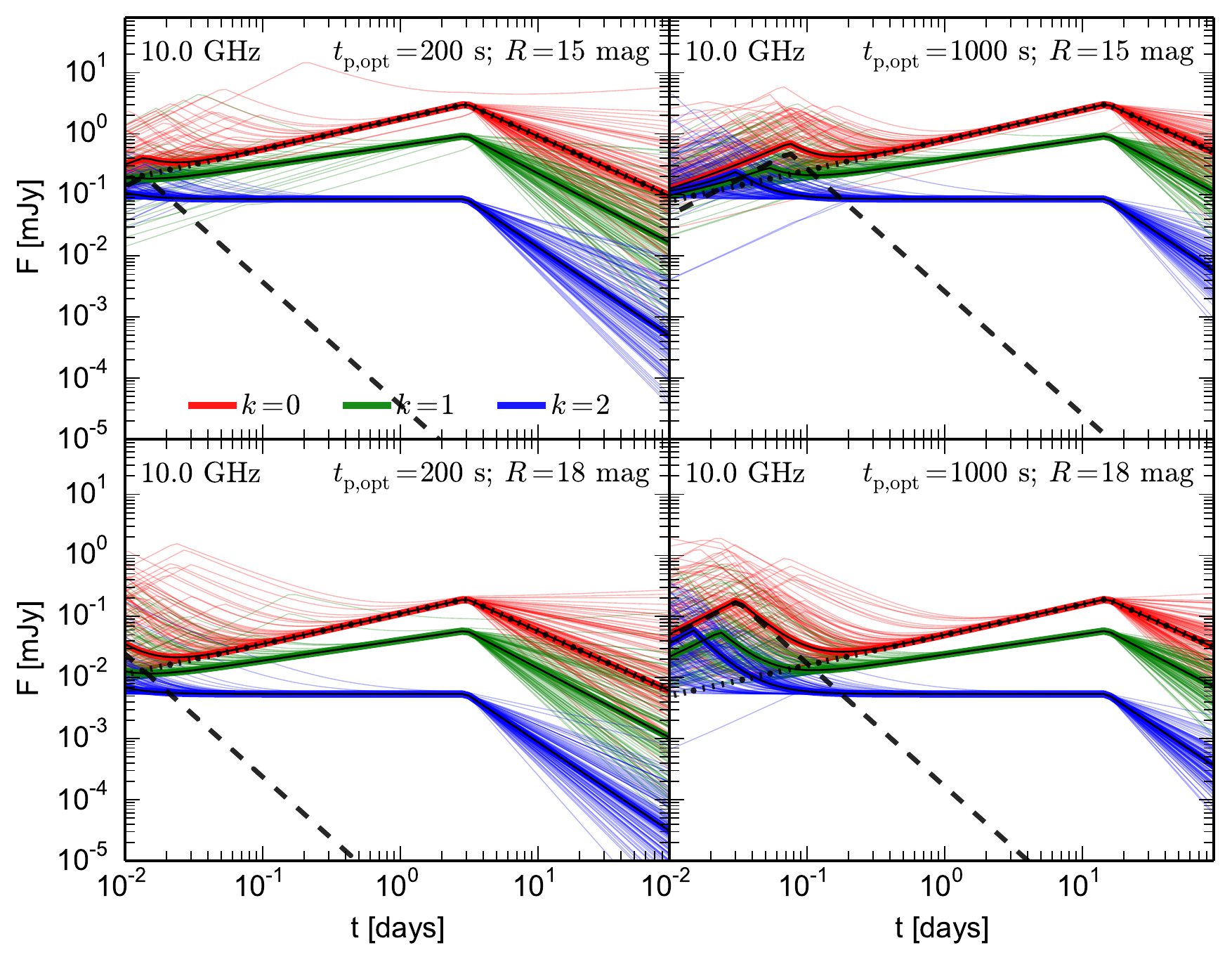}  
\includegraphics[width=0.82\linewidth]{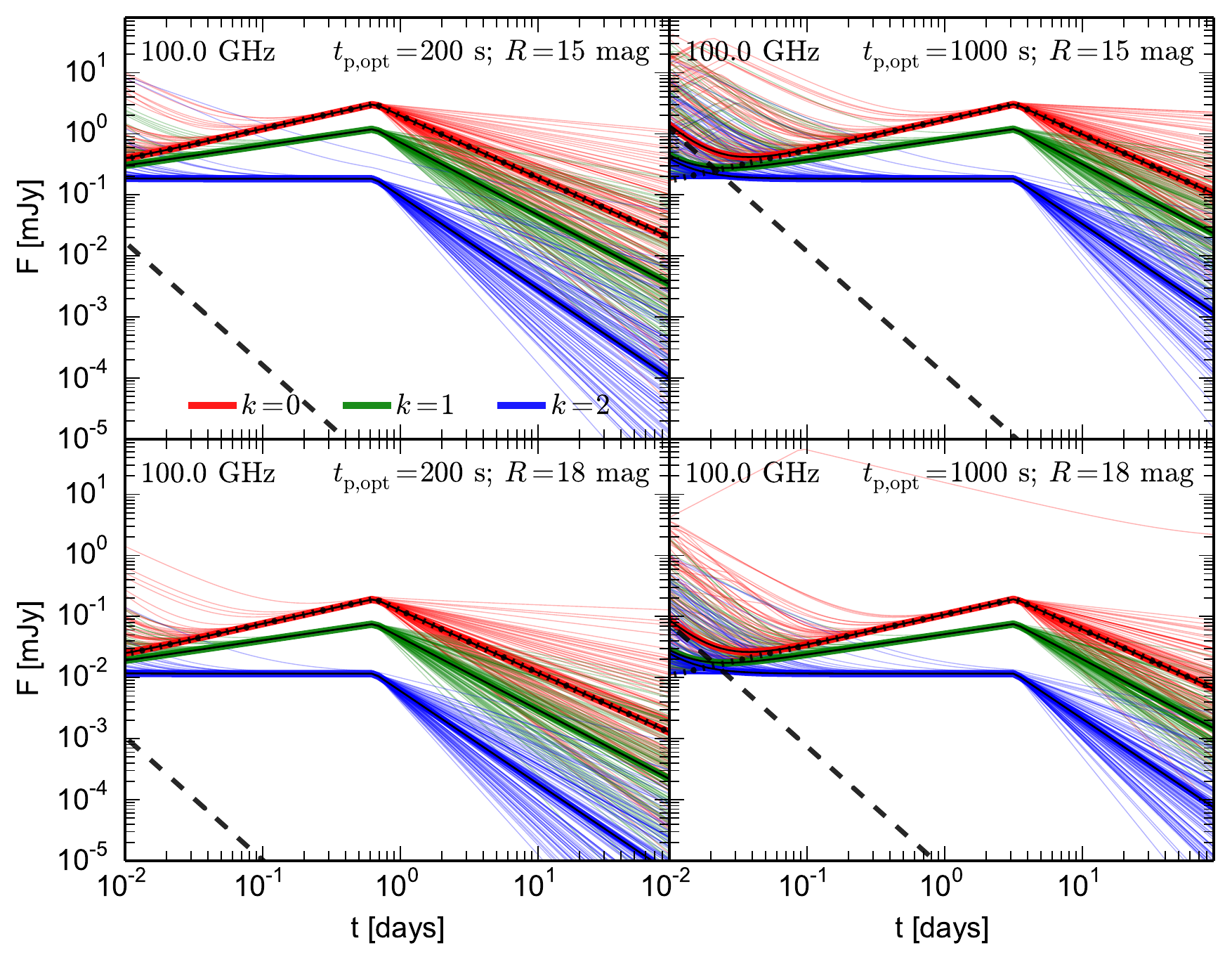}  
\caption{Predicated light curves for 10 GHz (top 4 panels) and 100 GHz (bottom 4 panels). See description in Figure \ref{k0lc}.}
\label{k1lc}
\end{center}
\end{figure*}

Figures \ref{k0lc} and \ref{k1lc} summarize the light curve predictions using the low-frequency model framework detailed in Section \ref{sect:model}. Figures represent three different observed radio frequencies ($1.4$ GHz, $10$ GHz, and $100$ GHz). Four panels of each frequency represent simulated light curves at three different circumburst medium structures $k$, for different combinations of optical peak time and brightness. In the following we consider some trends that emerge from the collection of light curves.

In general, it is evident that the need for rapid response at low frequencies is extremely important for RS physics, similar to optical wavelengths. The brightest and most evident RS emission occurs, in most cases, before 0.1 day, but for some cases can extend to the order of 1 day after the trigger, depending on observed radio frequency and circumburst medium structure $k$. A variety of factors contribute to conditions that lead to strong RS emission. 

\subsection{Optical peak properties ($t_{\rm p,opt}$, $\fpfs$, $\nufs$)}
\label{sect:optprop}

As the deceleration time increases, $\Gamma_0$ decreases, causing the brightness of the RS emission to decrease, but also causing the RS and FS peaks to occur later in time. Although seemingly counterintuitive, a later deceleration time allows for the possibility of catching the decay of the RS before it becomes surpassed by the increasing FS emission, thus helping to separate the two peaks in time, giving each better definition.

The brightness of the optical peak, and therefore of the FS peak, has a rather small overall effect in this model, basically just scaling the predicted radio brightness. We considered two cases: a `bright' optical peak of 15 magnitudes and a `dim' peak of 18 magnitudes. We find that a comparatively dim peak, still observable by existing optical facilities, can make the early-time radio RS emission slightly more observable, in the sense that it lowers the contribution of the FS emission while still maintaining a detectable flux density at low-frequency. 

The optical peak can be caused by the deceleration of GRB outflow or by the passage of the typical frequency through the optical band. A peak due to the former is expected to have a sharp rise, while a peak due to the latter would have a slow rise ($F \propto t^{1/2}$ for $k=0$ and shallower for $k>0$). If the typical frequency is above the optical band at the deceleration time, the optical flux is expected to rise as $t^{1/2}$ or shallower until the typical frequency passes through the optical band. Since such a slow rise is hardly observed (e.g. \citealt{melandri10}), we assume that the typical frequency $\nufs$ is already below the optical band $\nu_{\rm opt}$ at the deceleration time: $\chi < 1$. Changes when $\chi$ is significantly below $1$ are presented in Figure \ref{fig:param_change} (left), where we show the evolution of light curves for a case of $k=0$, $t_\mathrm{p,opt} = 1000\,\mathrm{s}$ and $R=18\,\mathrm{mag}$. The peak times of both FS and RS emission are shifted to earlier times, while peak fluxes are shifted to higher values, but especially in the RS regime flux density evolution is masked by the SSA limit, for which the break also shifts due to $\nurs$ shift. Lower $\chi$ thus makes RS emission less pronounced.

\subsection{Circumburst medium}
\label{sect:ism}

Different values of circumburst medium stratification $k$ affect the predicted light curves similarly to decreasing the optical peak flux. As shown in Figures \ref{k0lc} and \ref{k1lc}, the FS peak becomes less pronounced as we move from $k=0$ to $k=2$, due to the $-k/(2(4-k))$ dependence of the peak flux. The slope before the FS peak, however, becomes more shallow as $k$ increases, contaminating the early-time RS emission.

\begin{figure*}[!ht]
\begin{center}
\includegraphics[width=0.49\linewidth]{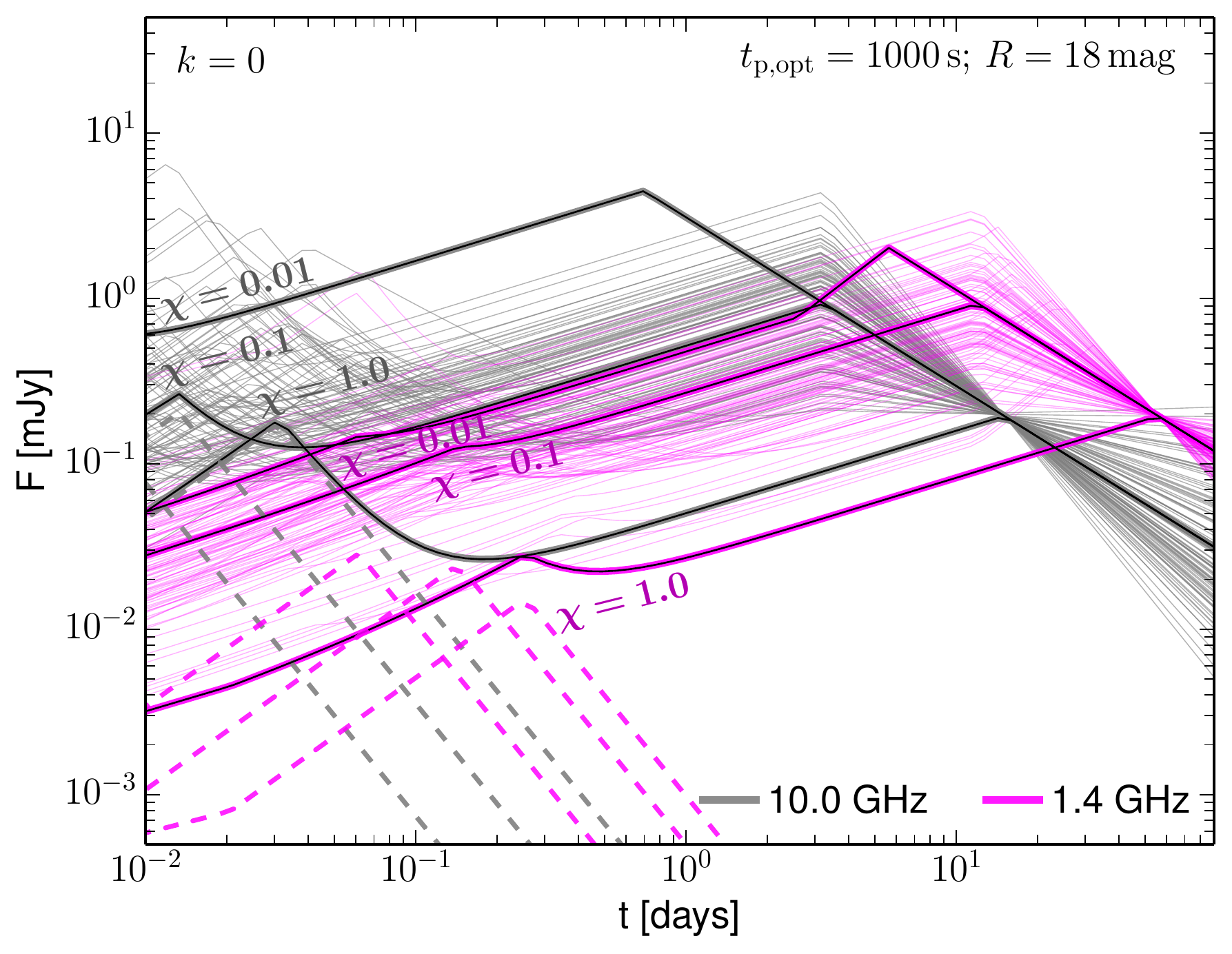}
\includegraphics[width=0.49\linewidth]{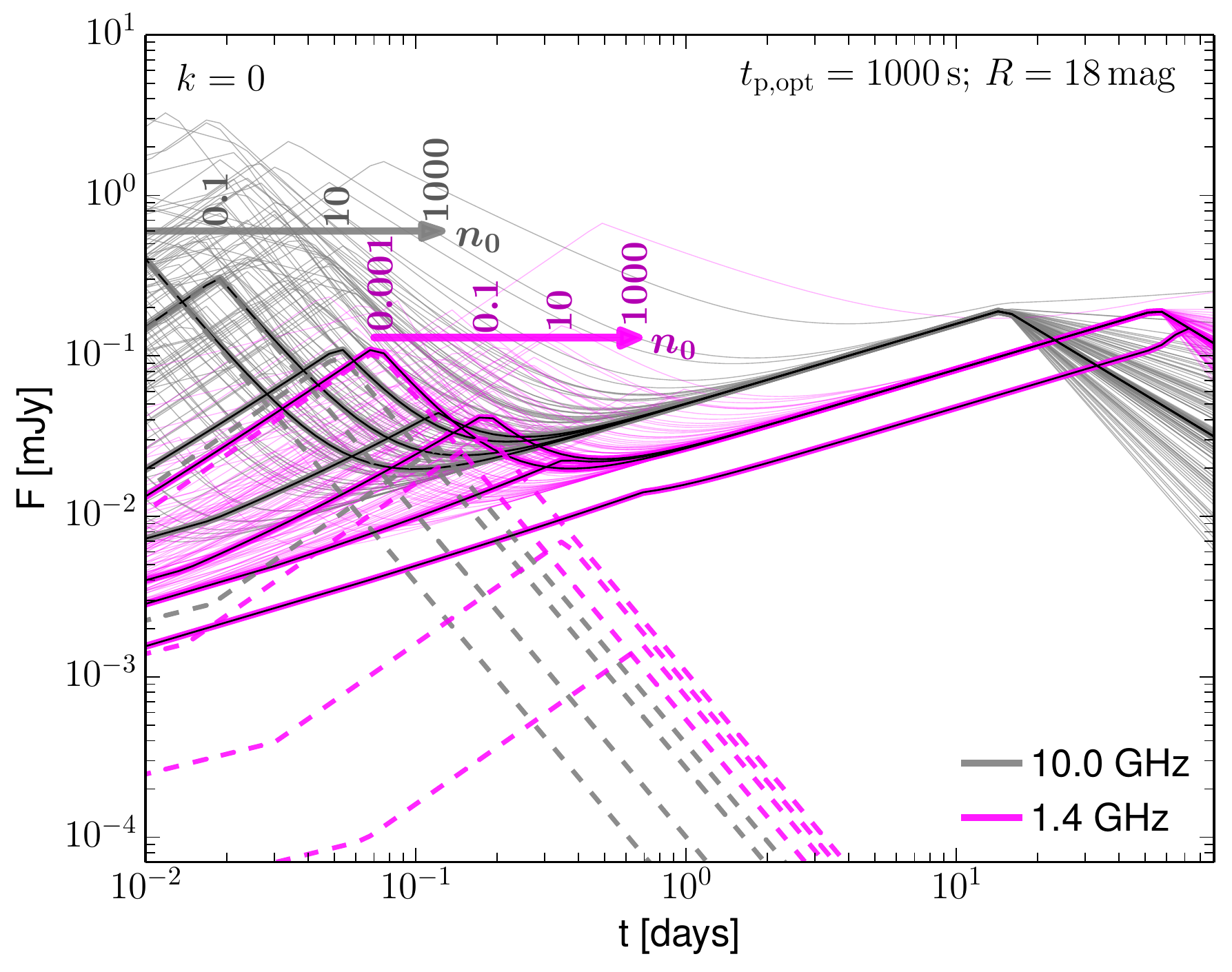} 
\caption{Evolution of light curves ($k=0$, $t_\mathrm{p,opt} = 1000\,\mathrm{s}$, $R=18\,\mathrm{mag}$ case) at 10 GHz (gray) and 1.4 GHz (magenta) when changing: (left) $\chi = \nufs/\nu_{\rm opt}$ parameter ($\chi = 1$, $0.1$, or $0.01$, from bottom to top). Smaller $\chi$ shifts FS and RS peaks towards earlier times, while peak fluxes are shifted to higher values but are masked by the SSA limit especially for the RS emission (dashed lines). Simulated light curves (thin lines) are only plotted for $\chi = 0.1$ case. (right) Circumburst medium density ($n_0 = 0.001$, $0.1$, $10$, or $1000$ $\mathrm{cm^{-3}}$, from left to right corresponding to RS peaks; for $n=1\,\mathrm{cm^{-3}}$ case see Figures \ref{k0lc} and \ref{k1lc}). Larger $n_0$ lowers $\Gamma_0$ estimate, shifting the RS peak towards later times and lower flux densities (dashed lines). Simulated light curves (thin lines) are only plotted for $n_0 = 0.1\,\mathrm{cm^{-3}}$ case.}
\label{fig:param_change}
\end{center}
\end{figure*}

Circumburst medium density $n_0$ determines the Lorentz factor estimate (Equation \ref{eq:gamma}) and consequently affects RS emission properties (Equations \ref{eq:nu_rs} and \ref{eq:fp_rs}) and the SSA limit approximation (Equation \ref{eq:ssa}). Due to the broad distribution of $n_0$ values obtained from the modeling of afterglows \citep{yost03,cenko11,japelj14}, which could span over 6 orders of magnitude, we did not include its parametrization in our simulations as otherwise the spread of the simulated light curves would be too wide to identify key trends and effects. However, we do assess the evolution of light curves when changing $n_0$ from $0.001\,\mathrm{cm^{-3}}$ to $1000\,\mathrm{cm^{-3}}$, as obtained from \citet{cenko11}. The resulting light curves are plotted in Figure \ref{fig:param_change} (right), for a case of $k=0$, $t_\mathrm{p,opt} = 1000\,\mathrm{s}$ and $R=18\,\mathrm{mag}$. We infer that in this case, the RS peak is much more pronounced in environments with lower circumburst medium density, while shifted towards earlier times. Same results also hold for different values of circumburst medium structure $k$.

\subsection{Self-absorption and $\nu_{\rm radio,obs}$}

The observed radio frequency has the two-fold effect of normalizing the level of the SSA flux and influencing the peak time of both RS and FS. The effect is more apparent in the latter as it is dependent on the time evolution of $\nufs$, while the RS will peak comparatively early due to its $\Gamma_0^2$ dependence. Around $\sim 1\,\mathrm{GHz}$, SSA tends to dominate and severely reduces the RS flux level, in some cases even removing any temporal structure that would help identify the RS emission. Since the SSA flux limit is $\propto \nu^2$, the effect is reduced with increasing observing frequency, aiding the observation of the RS at $\gtrsim 10$ GHz. However, when deciding on an observing strategy, it is necessary to balance these opposing effects by choosing a frequency that will not be too self-absorbed nor too high; in the latter case the RS and FS will peak very early and the RS signature would be very difficult to detect. 

\subsection{Numerical corrections}
\label{sect:discnum}

Further folded into our simulations are the changes in the light curves due to the addition of the numerical correction factors in Equations (\ref{eq:gamma}) - (\ref{eq:fp_rs_scaling}). Particularly at early time, many of the figures show bands of predicted light curves covering several orders of magnitude in flux density. This effect originates partly from the wide distribution of $\epsilon _e$ and its effect on the SSA limit but also from the numerical constants. We introduced these factors by assigning each randomly generated light curve a value for the dimensionless RS shell thickness parameter, $\xi _0$, from the observed distributions (Figure \ref{fig:xi} for $k=1$ case). The effects in intermediate ($\xi _0 \sim 1$) and thin-shell ($\xi _0 > 1$) regions are particularly pronounced and on the scale of 2 orders of magnitude (see Figure 3 in \citealt{harrison13} for $k=0$ case). In practice, this effect can be mitigated since $\xi _0$ can be approximated from the deceleration time and burst duration (Equation \ref{eq:xi0}). This approximation, however, assumes that the emitted shell has a homogeneous structure and that the deceleration time occurs after the end of the prompt emission. These are reasonable assumptions but may not always be valid, such as in GRB~061126 \citep{gomboc08}. This adds further uncertainty to the low-$\xi _0$ value tail of the distribution.

\begin{figure*}[!ht]
\begin{center}
\includegraphics[width=0.49\linewidth]{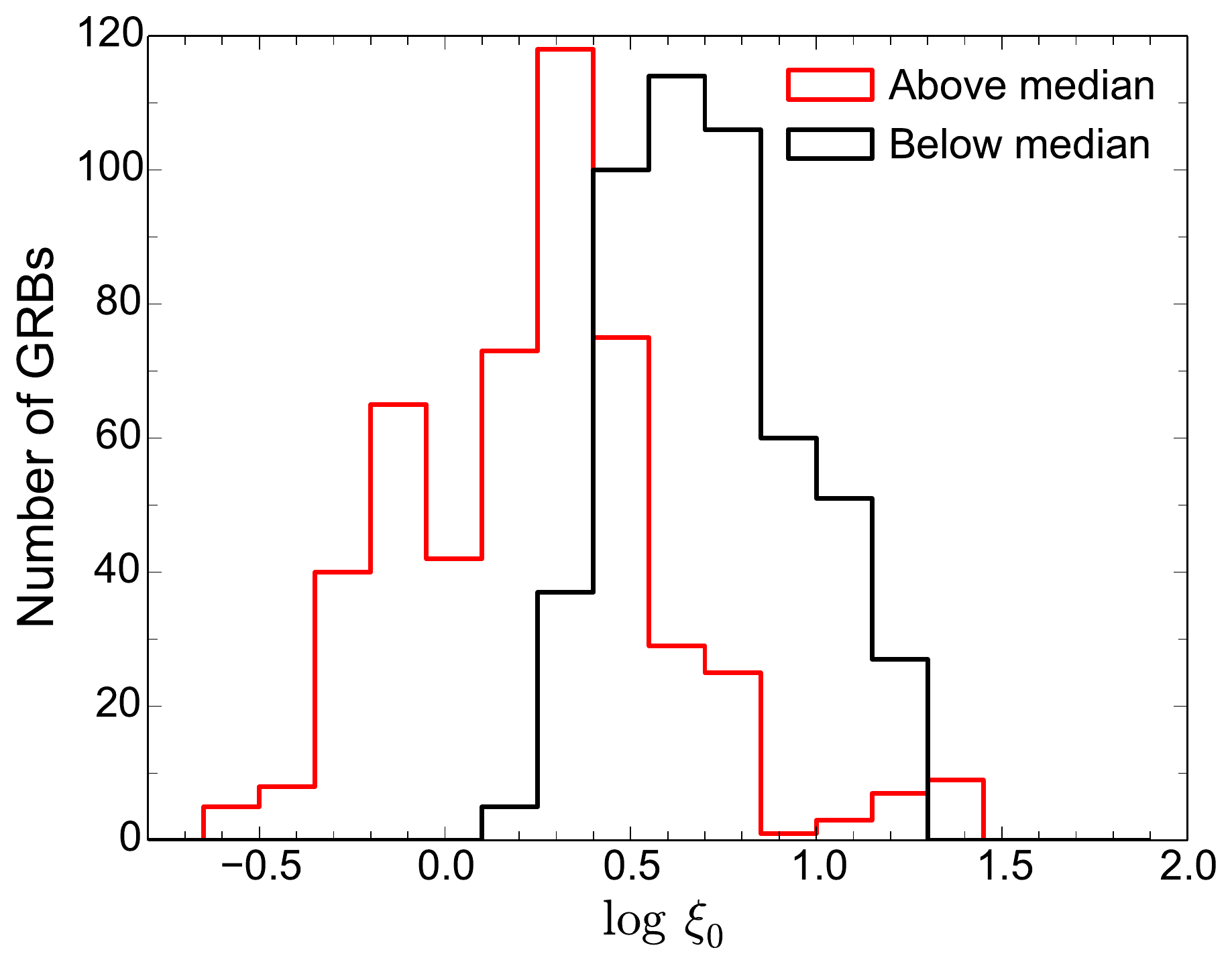}
\includegraphics[width=0.49\linewidth]{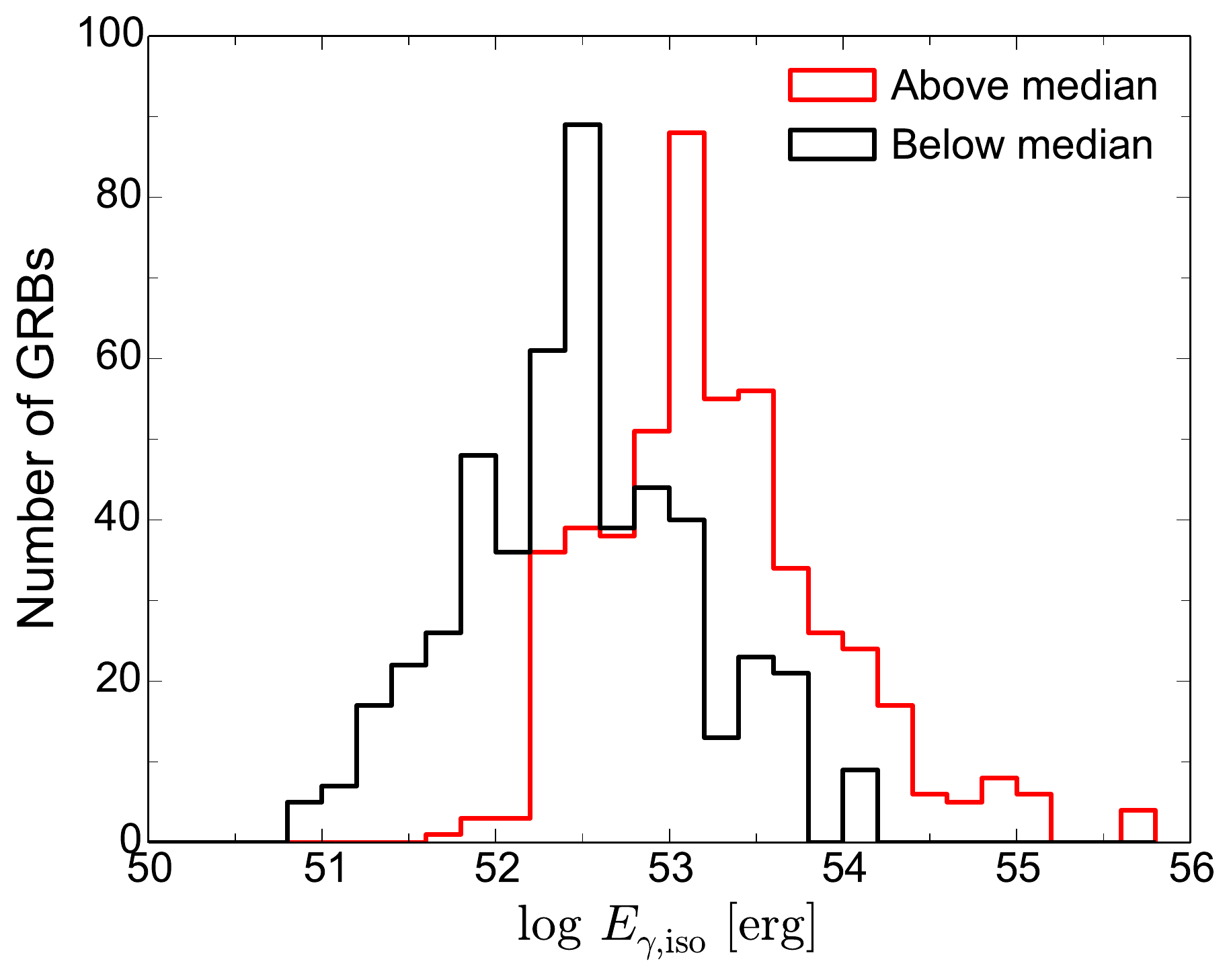} 
\caption{Distribution of $\xi _0$ (left) and $E_{\gamma \rm ,iso}$ (right) for a trial simulation ($t_{\rm p,opt}=1000\,\mathrm{s}$, peak mag $= 18\,\mathrm{mag}$, observed frequency = 10 GHz, $k=0$) for fluxes above and below the median. Bursts with above-median fluxes in the radio have larger energies and smaller $\xi _0$ (closer to thick-shell regime), helping to mitigate the effects of the numerical corrections.}
\label{xi_effects}
\end{center}
\end{figure*}

The broad $\xi _0$ distribution causes a large spread of RS light curves but also provides parameter spaces in which the RS can more easily be observed. We can better understand the effect that these numerical factors have by looking at the values of burst parameters that produce bright (dim), or above (below) median flux light curves in our simulations. Figure \ref{xi_effects} shows the $\xi _0$ and $E_{\rm \gamma,iso}$ distributions for an example trial of 10 GHz light curves from an assumed 18$^{\rm th}$ magnitude optical peak occurring at $1000$\,s after GRB. The brighter light curves are created by a combination of higher $E_{\rm \gamma,iso}$ and smaller $\xi _0$, since the former contributes to a larger $\Gamma_0$ and the latter's effect is less pronounced closer to the thick-shell regime.

\section{Discussion}
\label{sect:discussion}

Our simulations show that later optical peaks, assuming they indicate the fireball deceleration time, frequencies between $\gtrsim 1$ and $\lesssim 100$ GHz, and bursts that have a circumburst stratification closer to $k=0$, are preferred candidates for observing RS emission. Brightness of the optical peak has less of an effect in this model, although this is also dependent on the burst parameters and observing frequency, as the RS component may be partially self-absorbed or mixed with emission from the FS. RS emission with median flux levels of the order of $\mu$Jy-mJys should be visible at early times ($< 0.1 - 1\,\mathrm{day}$) with respect to the GRB trigger and are observable with currently operating facilities. Future facilities will enable shorter exposures times due to improved sensitivity, allowing for better light curve sampling.

As an example, with a standard array configuration (34 1-m, 9 7-m, and 2 TP antennas) observing at 100 GHz (ALMA Band 3), ALMA can detect afterglow emission (3$\sigma$) to limits of 10 (1, 0.1, 0.05) mJy with 0.06\,s (5.84\,s, 9.7\,min, 37.4\,min) integrations\footnote{Calculated using the ALMA sensitivity calculator found at http://almascience.eso.org/proposing/sensitivity-calculator}. Modest integration times can lead to the detection of many of the brightest afterglows, and in the $k=0$ and $k=1$ cases, a larger fraction of the dimmer afterglows. Longer wavelengths that are more affected by self-absorption will pose bigger problems, but should be observable in the extreme cases. In addition to sensitivity, facility response time is extremely important and capturing clear RS emission will require aggressive ToO campaigns.

Polarization can provide further evidence of RS emission. This technique has been pioneered at optical wavelengths and is rapidly becoming feasible at radio and sub-mm wavelengths. Polarization of GRB emission is a prediction of various theoretical models and has been observed in prompt $\gamma$-ray emission \citep{gotz09, yonetoku12, gotz14} and in optical to a level of 10-30\% in the first minutes after the burst \citep{steele09,uehara12,mundell13}, probing the fundamental fireball magnetic field. 

At later times (hours to days post burst), low levels of optical polarization of a few $\%$ have been reported by various works \citep{covino99,greiner03,wiersema12,wiersema14} and are thought to be due to FS-dominated emission, probing the shocked ISM. Radio polarization for a GRB has yet to be detected and upper limits of a few percent have been reported for a handful of bursts \citep{taylor04,taylor05,granot05} at days after the GRB trigger. Although emission at such late times most often comes from the FS, the underlying component of RS emission would help us understand its effect on the overall light curve behavior and polarization properties. Recently, \citet{vanderhorst14} undertook radio polarization observations of GRB~130427A at 1.5 and 2.5 days, finding similar upper limits despite the fact that these observations were taken at the radio peak time, which was associated with the RS emission. This event further stresses the importance of rapid multi-wavelength observations if we are to understand the size and scale of the magnetic fields in GRB ejecta.

Looking beyond current capabilities, the SKA will provide unprecedented sky coverage at frequencies around 1.4 GHz (e.g., \citealt{ghirlanda13,burlon15}) and has the potential to observe the very early-time RS signature of GRB emission. However, at such low frequencies, the early time flux in many cases will mostly be suppressed by self-absorption (see Figure \ref{k0lc}) and expected RS flux densities at $0.1$ days after the GRB trigger are on average lower than $\lesssim 10\,\mathrm{\mu Jy}$. But depending on burst's parameters, it is still possible in some cases to detect distinguishable RS peaks with peak flux up to $\sim 0.1\,\mathrm{mJy}$, as indicated especially by thin lines from simulated light curves (especially for `dim' and `late' optical peaks occurring in $k=0$ type medium, see Figure \ref{k0lc}). Such fluxes could be achieved by the SKA with a reasonable ($\sim 10-30$ minutes) integration times with a typical array setup \citep{ghirlanda13}. This exposure time would be short enough to achieve a reasonably deep limit without compromising valuable temporal resolution.

\section{Conclusions}
We have presented a collection of predicted radio afterglow light curves in the framework of the GRB standard model, considering contributions from both FS and RS emission within an updated low-frequency model framework (\citealt{mundell07,melandri10}). Our results show that:
\begin{itemize}
\item Inclusion of correction factors from numerical simulations that parametrize shell-thickness regime (\citealt{harrison13}; 2015 in prep) provides more accurate estimates of RS emission.
\item RS emission can significantly alter the temporal behavior at early times, producing a distinct ``radio flare'' (e.g. \citealt{kulkarni99}). The brightness and shape depend on various parameters, but are mostly affected by the SSA, which is stronger at lower frequencies. 
\item RS emission could be best identified at frequencies around $\sim 10$ GHz, peaking at $\lesssim 0.1$ day after the burst, depending on burst's parameters. At frequencies around $\sim 1$ GHz, RS emission is more likely to be strongly suppressed by the SSA, while at higher frequencies (around $\sim 100$ GHz), RS emission tends to peak at earlier times ($\lesssim 0.01$ day after the burst). 
\item Current and future radio facilities with high sensitivity and short response times could detect radio flares with peak flux densities around $F \sim 0.01-0.1\,\mathrm{mJy}$ ($F \sim 0.1-10\,\mathrm{mJy}$) at $t \lesssim 1$ day ($t \lesssim 0.1$ day) after the burst, at $1.4$ GHz ($10$ GHz), respectively. 
\end{itemize}

\acknowledgments

We thank the anonymous referee for valuable comments and suggestions which improved the paper. D.K. and F.J.V. acknowledge support from the UK Science and Technology Facilities Council. C.G.M. acknowledges funding from the Royal Society, the Wolfson Foundation, and the UK Science and Technology Facilities Council. A.M. acknowledges funding from ASI grant INAF I/004/11/1.

\end{document}